\documentclass[eprint,review=false
]{actapoly}

\usepackage[utf8]{inputenc}

\begin{document}

\title[Gamma-Ray Bursts and Optical Brightness: Insights from Swift Satellite]{Gamma-Ray Bursts and Optical Brightness: Insights from Swift Satellite}

\correspondingauthor[I. I. Racz]{Istvan I. Racz}{my}{racz.istvan@uni-nke.hu}
\author[Y. Nehme]{Yasmin Nehme}{their}
\author[L. G. Balazs]{Lajos G. Balazs}{their,third}

\institution{my}{University of Public Service, Department of Natural Sciences, 2 Ludovika tér, H-1083 -- Budapest, Hungary}
\institution{their}{Eötvös Loránd University, Faculty of Science, Institute of Physics and Astronomy, Department of Astronomy, Pázmány Péter sétány 1/A., H-1117, Budapest, Hungary}
\institution{third}{HUN-REN Research Centre for Astronomy and Earth Sciences, Konkoly Thege Miklós Astronomical Institute, Konkoly-Thege Miklós út 15-17., H-1121, Budapest, Hungary}

\begin{abstract}
During the examination of optical properties of gamma-ray bursts (GRBs) measured during the life of the Neil Gehrels Swift Observatory satellite, significant discoveries were made. While the satellite measures gamma, X-ray, and optical data simultaneously, in many cases, only an upper limit is established in the optical domain. Hence, survival analysis, a.k.a. dependability modelling, was necessary, serving as a tool for studying samples where some cases are bound only by the upper limit, as mentioned earlier. This method is used to analyse datasets that may have upper or lower bounds rather than precise observations, but nevertheless contain relevant information.
It was previously established that the duration, gamma fluence, and peak flux all have a substantial effect on the optical brightness distribution, probably due to the energy of the beam from the GRB's central engine.
However, after analysing more than 200 data points, we no longer see this association. This disagreement shows that the previously reported impact may not be as significant as previously derived, needing more research to better understand the underlying processes that determine the optical brightness in connection to gamma-ray properties.

\end{abstract}

\keywords{data analysis -- gamma-ray bursts}

\maketitle

\section{Introduction}

Gamma-ray bursts (GRBs) are the most energetic and luminous events observed in the Universe since the Big Bang \cite{meszaros_2006,Kumar}. First discovered in the late 1960s, GRBs are brief, intense flashes of gamma-ray radiation that can last from a few milliseconds to several minutes \cite{1973ApJ...185L..57W}. These bursts originate from distant galaxies, often billions of light-years away, making them valuable probes of the early Universe.

The fundamental feature of GRBs is their tremendous energy output, with some bursts releasing more energy in a few seconds than the Sun will release in its entire 10-billion-year lifetime. GRBs are often classified into two types based on their duration: short GRBs, which last less than two seconds, and long GRBs, which can range from two seconds to several minutes. According to the dominant hypotheses, short GRBs occur from neutron star mergers \cite{ns_merging}, whereas long GRBs are caused by the collapse of massive stars into black holes (collapsar model) \cite{collapsar, meszaros_fireball}. The connection with gravitational waves, which has already been observed several times, can serve as evidence for the first case \cite{2017PhRvL.119n1101A, veres_gw170814,2016A&A...593L..10B,2017CoSka..47...76B,2018Ap&SS.363...53H}. In addition to the two main models, other types of GRBs were discovered in the 1990s and early 2000s \cite{hoi1998} with lengths of a few seconds, i.e. of medium length, and spectral hardness different from the two above \citep{hor04,hoi2006,2020MNRAS.494.4343K}. The physical model of this intermediate group could not be perfectly determined but it seems that the X-ray flash events could have a connection to them \citep{hoi2010, 2010ApJ...725.1955V, pinter_2018, Xiongwei2018}. 
In addition to their initial gamma-ray emission, GRBs frequently produce an afterglow that glows at various wavelengths, including X-rays, ultraviolet, optical, and radio waves, providing critical information on the environment and processes surrounding these explosive events \cite{zhang_2018}.

\subsection{Large structures}

In\citet{balazs1998}, the authors have demonstrated  that the angular distribution of short GRBs is not entirely isotropic \cite{, ,2000ApJ...539...98M,2000A&A...354....1M,2009BaltA..18..293M,2009A&A...493...51B}, in contrast to early observations and models of the sky distribution of GRBs. In fact, there is mounting evidence that GRBs exhibit notable, giant anisotropies \cite{1999A&AS..138..417B,2008MNRAS.391.1741V,2010A&A...510A.105P,2018ApJ...855..101H,2019Ap&SS.364..105H,2019MNRAS.486.4823T,2020MNRAS.498.2544H}.

\begin{figure}
\centering
\includegraphics[width=\linewidth, trim={10 5 10 5},clip]{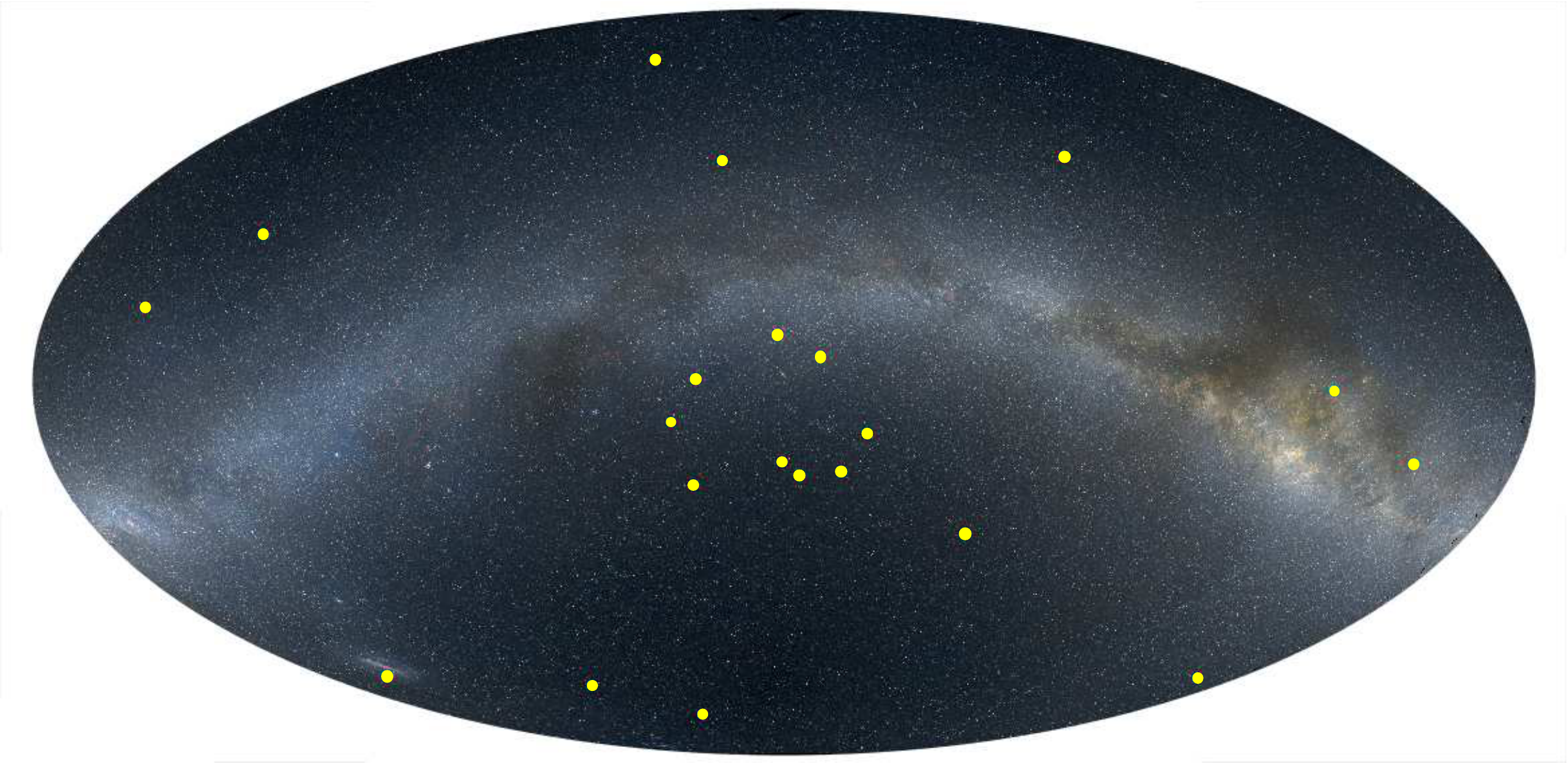} 
\\[3mm]
\includegraphics[width=\linewidth]{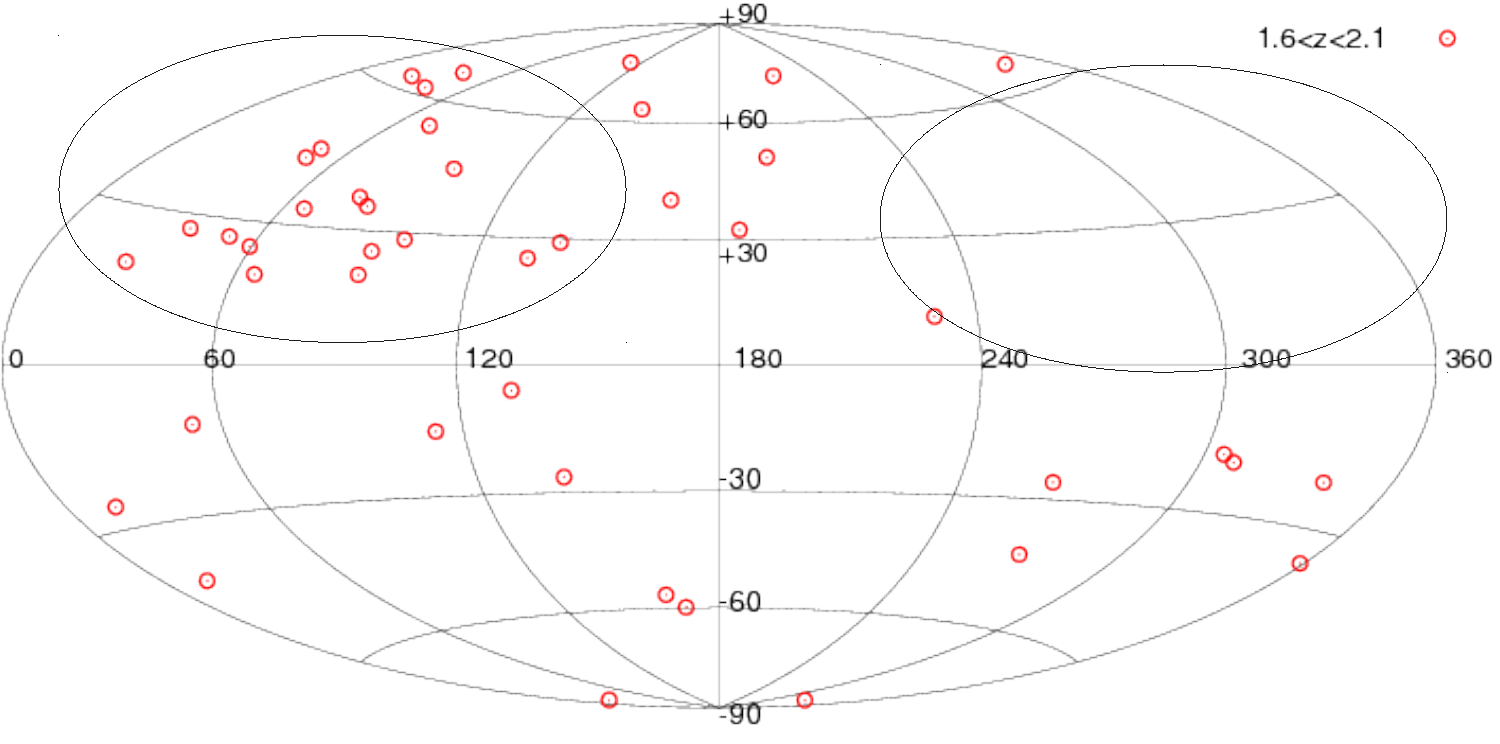} 
\caption{The image above shows the Great Ring-like GRB cluster, while the image below shows the Corona Borealis Great Wall.}
\label{fig:grbstructures}
\end{figure}

Insight into the vast and intricate architecture of the Universe has recently been gained through the discovery of massive cosmic structures.
\cite{2000A&A...354....1M, 2000ApJ...539...98M, 2022Univ....8..221H}. A noteworthy finding among them is the GRB ring, a massive ring-like configuration of gamma-ray bursts extending over 5 billion light-years that complicates our notion of large-scale cosmic homogeneity \cite{ring_1,ring_2}. The Hercules-Corona Borealis Great Wall is another noteworthy structure \cite{great_wall_1}. At over 10 billion light-years in length \cite{great_wall_2}, it is the largest known structure in the Universe (see Figure~\ref{fig:grbstructures}). This massive GRB supercluster leads to new questions about the formation and evolution of the Universe, defying previous expectations of the size limits for cosmic structures. It is possible that the distribution of matter in the Universe is more varied and complex than previously believed if these enormous structures are present, which has significant implications for cosmology \cite{1992Ap&SS.191..107P}. 

If we accept the widely held view that there are only two physical hypotheses for the origin of GRBs, and taking the hypernova concept as a starting point, we can state the following: in the collapsar model, which occurs upon the collapse of a very massive star, it is obvious from the physics scenario that these stars are bound to their birthplace due to their short existence. When \cite{farthest_grb_spec} published the highest redshift of a spectroscopic GRB, it was the Universe's most distant object \citep{2019AN....340..681B}. We believe that the distribution of GRB redshifts, which reaches at least $z\approx10$, and their correlation with the explosive death of massive stars, or GRBs, can make them a special and potent tool for cosmology \citep{2009Natur.461.1258S,2021ExA....52..219T}. Finding hundreds of GRBs, including those with high redshift ($z>6$) \citep{swom, theseus01}, will be one of the main goals of two upcoming space missions, Space Variable Objects Monitor (SVOM) and Transient High-Energy Sky and Early Universe Surveyor (THESEUS) satellites, among others, in an effort to address these important questions. It also gives us insights into the evolution of stars in the early Universe, and the role of early galaxies where stars form.

\subsection{GRBs as tracers of high SFR}

Gamma-ray bursts are not just exotic cosmic occurrences, but also useful markers of galaxies' star formation rates \cite{2019MNRAS.489.3778D,2019PASJ...71...10T}. 

\begin{figure}
\centering
    \includegraphics[width=\linewidth]{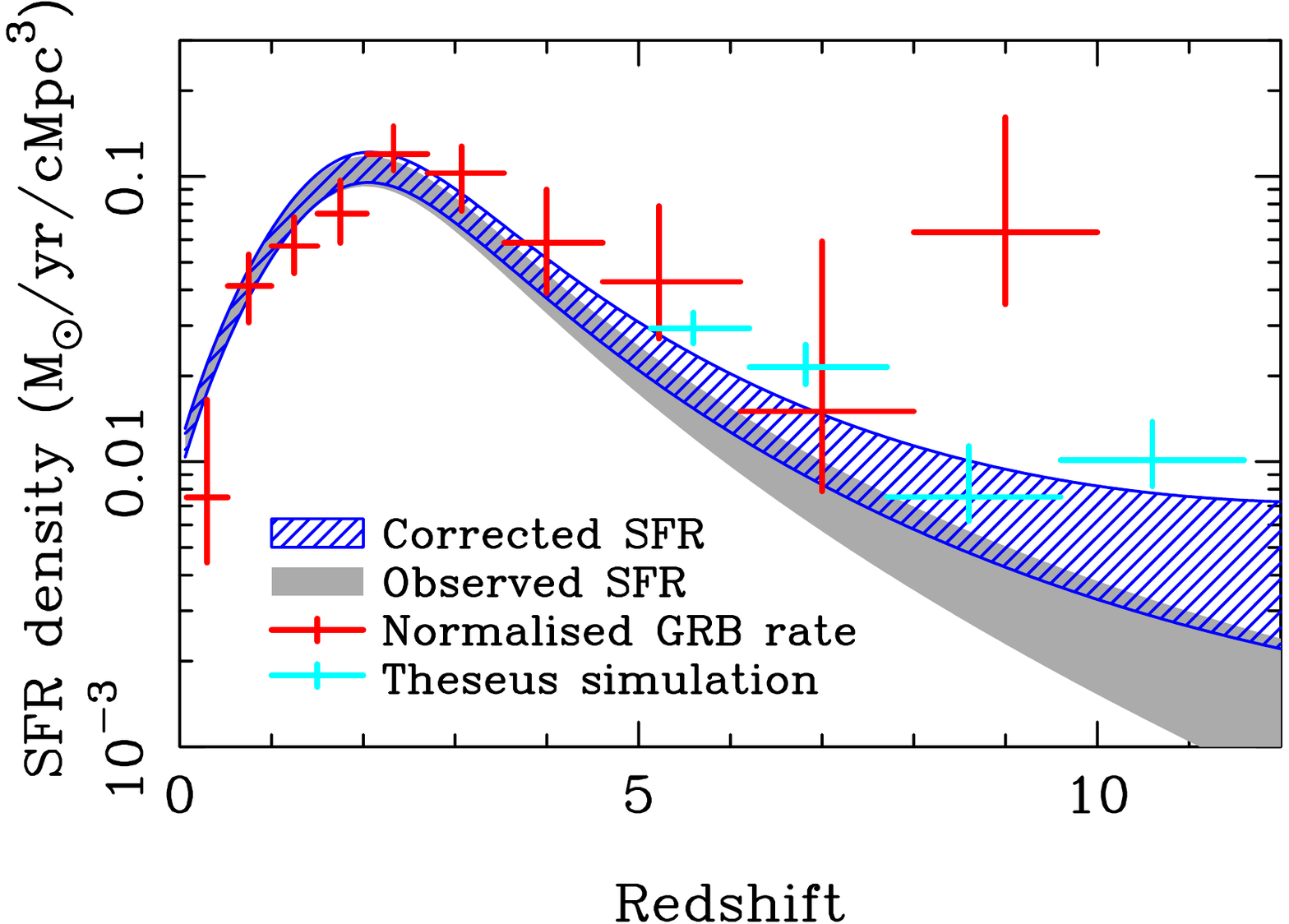} 
\caption{Source: Tanvir et al.(2021) \citet{2021ExA....52..219T} Fig 2. Star Formation Rate density from GRBs based on various assumptions (GRB rate to SFR ratio, GRB progenitor metallicity) and as a function of redshift (shaded curve based on the observed galaxy population, blue hatched region accounts for galaxies below the detection limit). Current constraints from the available GRB sample are shown by red points. Corresponding estimates from a representative THESEUS sample are indicated by cyan points. \label{fig:theseus_starforming}}
\end{figure}

It was shown already, using ISO observations (see eg. \cite{Stickel1998},  \cite{Heraudeau2004}), that the star forming rate density (SFRD) decreased in the last 5 billion years. While the SFRD curve is well understood in the $0 \leq z \leq 2.5$ interval (see eg. \cite{Madau2014}), we need more indicators of the star formation in the early Universe, and one possibility is using GRBs resulting from collapsars.
Because of their brief lifespan, they are limited to fruitful star formation areas. As a result, the frequency and distribution of GRBs can shed light on the locations and rates of star formation within galaxies. GRBs observed by mankind are found at significant cosmic distances, up to redshifts of 10, so astronomers can investigate GRBs to follow star formation activity in various epochs and settings of the Universe, gaining a better knowledge of how galaxies evolve and produce new stars over time (see Figure~\ref{fig:theseus_starforming}).

\section{Materials and methods}

\subsection{Data}

NASA launched the multi-wavelength \textit{Neil Gehrels Swift Observatory} telescope (formerly known as 'Swift' Space Telescope) in 2004 with the main goal of studying gamma-ray bursts (GRBs). Among the Universe's most powerful and explosive occurrences are these bursts. The three main instruments onboard Swift are the Ultraviolet/Optical Telescope (UVOT), the X-Ray Telescope (XRT), and the Burst Alert Telescope (BAT) \cite{swift_bat_2005, xrt_2005, uvot_2005}. The BAT's main task is to find GRBs and provide exact locations to the other sensors to point in a timely manner. When a GRB is discovered, the XRT is devoted to taking precise afterglow X-ray pictures and spectra, which enable researchers to examine the high-energy processes that follow the original burst. In the meantime, the UVOT monitors the afterglow in the visible and ultraviolet spectrum, offering further information that aids in deciphering the environmental and physical circumstances underlying these explosive occurrences. With this equipment, Swift is able to offer a detailed picture of GRBs at various wavelengths, which is greatly advancing our knowledge of these cosmic occurrences.

We used information from Swift's GRB observations from official public catalogue\footnote{\url{https://swift.gsfc.nasa.gov/archive/grb_table/}}. The catalogue includes most of the $\gamma$-ray parameters, as well as some X-ray and UV optical parameters and information on all the three Swift instruments. In this work, we have focused on the optical data recorded with the ultraviolet/optical telescope (UVOT).
There were 504 detected GRBs in total, 296 had an accurate measurement, 172 were censored and 36 with NA V mag. In some cases, the 11h X-ray flux value was not provided (26 cases), and the gamma spectral class was also missing in a few instances (5 cases).
Out of the 437 GRBs observed, 282 had precise measurements of their V magnitude, whereas for 155 events, only a limiting magnitude (lower limit) could be determined for various reasons.
Our goal was to identify the magnitudes affecting the optical brightness distribution during these occurrences. In order to do this, we examined a number of characteristics from the X-Ray Telescope (XRT) and the Burst Alert Telescope (BAT). For the BAT, we considered the photon index, 1-second peak photon flux, fluence, and T90 duration. From the XRT, we looked at the first temporal index, spectral index, early flux, 11-hour flux, and 24-hour flux. These characteristics gave us a large dataset that we could use to investigate the relationships and correlations that result from the GRB afterglow optical brightness.

\subsection{Survival analysis}
The study of the lifespan of different things is the focus of the statistical area of the survival analysis, a.k.a dependability modelling. Broadly speaking, we look at the moment at which an event, such as like the termination of a state, occurs \cite{Hacking2013-qv,2014styd.confE..99R}. The calculation of the so-called survival function and the associated significance tests form the basis of the methodology \cite{1985ApJ...293..192F, 1986ApJ...306..490I, 2012msma.book.....F}. When applied to biology, the function under examination shows how, based on a starting point or event, the estimated number of survivors should decline with time \cite{Andersen2005-ju}.

Here are some further examples of applications related to natural science:

\begin{enumerate}[label=(\alph*)]
\item Biology: By using survival analysis, one can get a greater understanding of the origins of illnesses, such as cancer, the lifespan of organisms, and the ways in which environmental or genetic variables influence an organism's ability to live.
\item Ecology: In ecological research, it can be used to monitor changes over time and to study individual survival rates and population extinctions.
\item Environmental science: It can be used to anticipate the effects of climate change and study the effects of environmental adjustments.
\item Geology: It can also be used to analyse volcanic eruptions, earthquakes, and other geological phenomena to determine the expected length or frequency of such occurrences.
\end{enumerate}

It is crucial to keep in mind that a study or research project has a certain amount of observation time. This can be due to a lack of resources (time, money, etc.) or just the researcher's desire to eventually publish the findings. Each research will therefore have a specific start and end date. 

It is also critical to remember that certain data have undergone interval, left, and right censorship. The right-hand limitation is applicable if, at the conclusion of the inquiry, the occurrence was not seen. In certain cases, the left-hand limit is reached when the event occurred prior to the start of the inquiry. This may be the case, for example, if we start looking at how a community's street lighting fixtures are operating and find out that some of them have already been damaged. It is feasible that some lamps will continue to work if, as in the previous case, the observation only lasts for a year. This is because we have both left and right constraints in time, which is how we generate our interval-censored data series.

In this astronomical endeavor, we know that an object's brightness is below the detection limit in a certain colour if our telescope was unable to detect it in that colour. This realisation connects to the topic of observational limitations and data censorship. Just as the researcher has to deal with the constraints of observation periods and the censorship of data in social research, astronomers are also restricted with limitations of their instruments. Therefore, the same principle and methods of survival analysis can be used in astronomy in order to achieve an accurate data analysis.

\subsubsection{The survival function}
The survival function $S(t)$, one of the most crucial quantitative terms in survival analysis, gives the probability that the subject of study will survive beyond a certain time $t$, meaning that the event of interest has not occurred yet. If we denote the survival time as $T$, a non-negative random variable that describes the time until the event of interest occurs, and $t$ as the exact point in time at which the survival probability is evaluated, then the survival function $S(t)$ can be mathematically defined as:
\begin{align}
S(t) = P(T > t)
\label{eq:01}
\end{align}
Note that $t$ and $T$ can refer to any variable of interest, not necessarily time (e.g. object's brightness).

From the definition of the survival function $S(t)$, it is clear that $S(t)$ is complementary to the cumulative distribution function $F(t)$ (CDF), the probability that the event has occurred by time $t$  $P(T\leq t)$, implying the following relation, 
\begin{align}
S(t) = 1 - F(t)
\label{eq:02}
\end{align}

It follows that the survival function $S(t)$ has the following properties:
\begin{enumerate}[label=(\alph*)]
\item $S(0) = 1 $, that is, the event has not occurred yet
\item $S(\infty)=0 $, that is, the event will eventually occur
\item $S(t)$ is non-increasing function, s.t. \\
$ S(t_1) \leq S(t_2), \text{ if } t_1 > t_2$

\end{enumerate}

There are several parametric, semi-parametric and non-parametric methods of estimating the survival function, such as exponential, Weibull, or log-normal\cite{survival-book}.  
One of the most frequent methods used, the Kaplan-Meier (KM) method, is a non-parametric estimator given by
\begin{align}
\hat{S}_{KM} (t) = \prod_{t_j \leq t} \left( 1 - \frac{d_j}{n_j} \right)
\label{eq:03}
\end{align}
where \begin{itemize}
  \item $d_j$ is the number of events at time $t_j$.
  \item $n_j$ is the number of subjects at risk\footnote{"at risk" refers to events that have yet to happen and are still under observation, i.e. not censored.} just before time $t_j$.
\end{itemize}

Visually, the Kaplan-Meier survival function is a step function that produces a statistical curve where at each point in "time", it shows the estimated survival percentage of subjects. Meaning, the curve changes value as a step down only when the desired event occurs and remains flat in intervals where no events occur. 

\subsubsection{The risk of the event}
Another important quantitative term to define in survival analysis is the hazard function $\lambda(t)$. Also known as the risk ratio, the hazard function provides a measure of the risk of the event occurring at a specific time $t$. That is, the hazard function $\lambda(t)$ is the probability that the event occurs per unit time at time $t$, given that we know the event still has not occurred. This can be shown by the following formula: 
\begin{align}
\lambda(t) = \lim_{\Delta t \to 0} \frac{P(t \leq T < t + \Delta t \mid T \geq t)}{\Delta t}
\label{eq:04}
\end{align}
Equation~\ref{eq:04} describes the slope of the cumulative survival curve given survival up to that time.

$\lambda(t)$ can now be written in terms of $S(t)$ as:
\begin{align}
\lambda(t) = \frac{f(t)}{S(t)}
\label{eq:05}
\end{align}
where $f(t)$ is the probability density function (PDF) defined as $f(t) =\frac{dF(t)}{dt}$.

The hazard function is always non-negative, $\lambda(t)\geq0$, and unlike $S(t)$, it has no upper bound. 

\subsubsection{Proportional hazards model}

In survival analysis, we often need to evaluate the effect of several explanatory variables on the time a specified event occurs. The Kaplan–Meier estimate falls short on this task as it only examines the effect of a single factor at a time. To solve this issue, the proportional hazards regression approach, also known as the Cox model, was created by Sir David Cox in 1972 \cite{Cox1972-uv,1986ApJ...306..490I}. The core idea of the Cox model is to estimate the hazard, or the event occurrence rate, at any given time point for an individual, given their covariates, relative to a baseline hazard function that is common to all individuals.

The process involves calculating the relative risk of the explanatory factors under investigation. In this context, a baseline group is always used as a risk-free comparison for the relationship, providing a reference point against which the effects of covariates are measured. This comparison is essential for interpreting the effects of the explanatory variables in terms of increased or decreased risk relative to the baseline.

The Cox model does not assume any particular baseline hazard function, allowing for flexibility in the shape of the hazard over time. This semi-parametric nature of the model makes it a robust tool for handling various types of survival data. The model can be expressed as:

\begin{align}
\lambda(t|X_i) = \lambda_0(t) \exp(\beta_1 X_{i1} + \cdots + \beta_p X_{ip})    
\label{eq:06}
\end{align}

where \( \lambda(t|X_i) \) represents the hazard function, or the instantaneous risk of the event occurring at time \( t \) for an individual with covariate vector \( X_i \). The term \( \lambda_0(t) \) denotes the baseline hazard function, which is the hazard for a baseline group that serves as the reference. The coefficients \( \beta_1, \ldots, \beta_p \) are the regression coefficients that quantify the effect of each covariate on the hazard.

This technique enables researchers to demonstrate whether a given treatment or intervention is effective by comparing the hazard rates between different groups while adjusting for other variables. For instance, in clinical studies, the Cox model can be used to evaluate the effectiveness of a new drug by comparing the time to recovery between the treatment and control groups while adjusting for potential confounders, such as age, sex, and comorbidities.

The Cox proportional hazard model has become a cornerstone in the analysis of survival data due to its ability to handle censored data and incorporate multiple covariates simultaneously, making it an invaluable tool in medical research, epidemiology, and various other fields.

\section{Results}
\subsection{Survival function of GRBs}

\begin{figure}
\centering
\includegraphics[width=\linewidth]{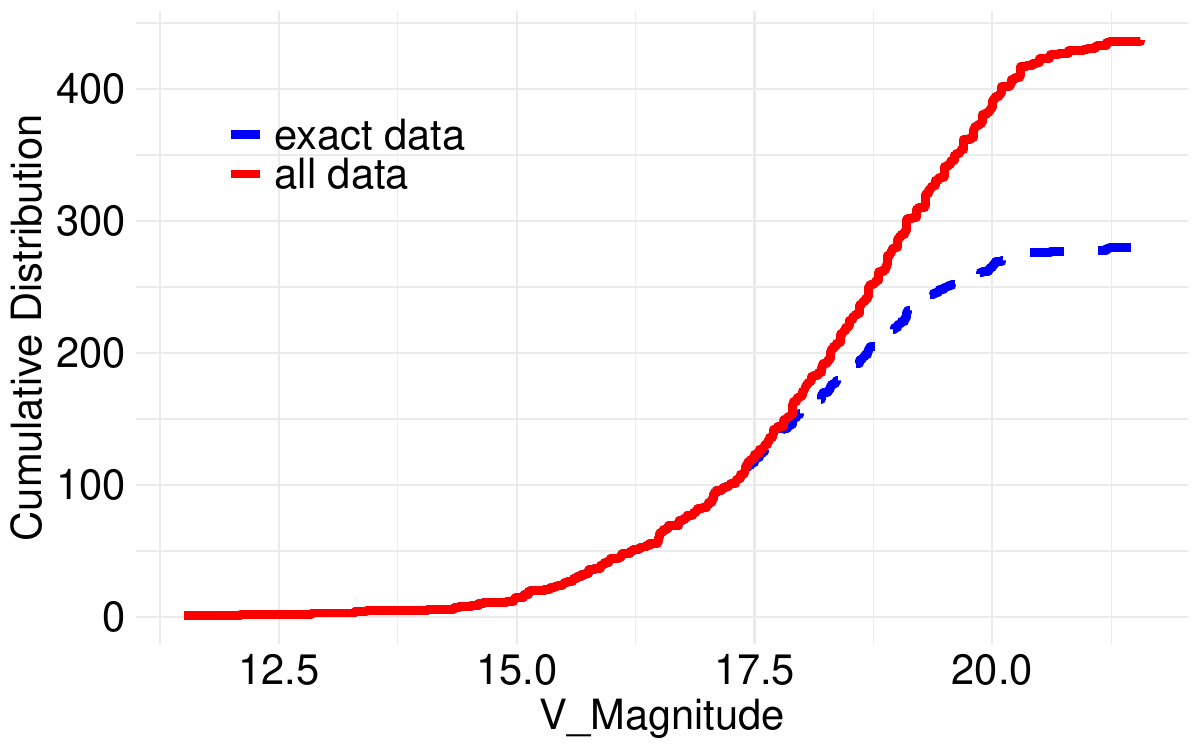} 
\caption{The figure shows the cumulative distribution of the Swift UVOT V band measurements. The red line shows the distribution of all data, the blue dashed line shows the distribution of only the exact measurements. You can clearly see how the censored measurements appear at lower magnitudes.}
\label{fig:result_survival_compre}
\end{figure}

\begin{figure}
\centering
\includegraphics[width=\linewidth, trim={2 10 0 10},clip]{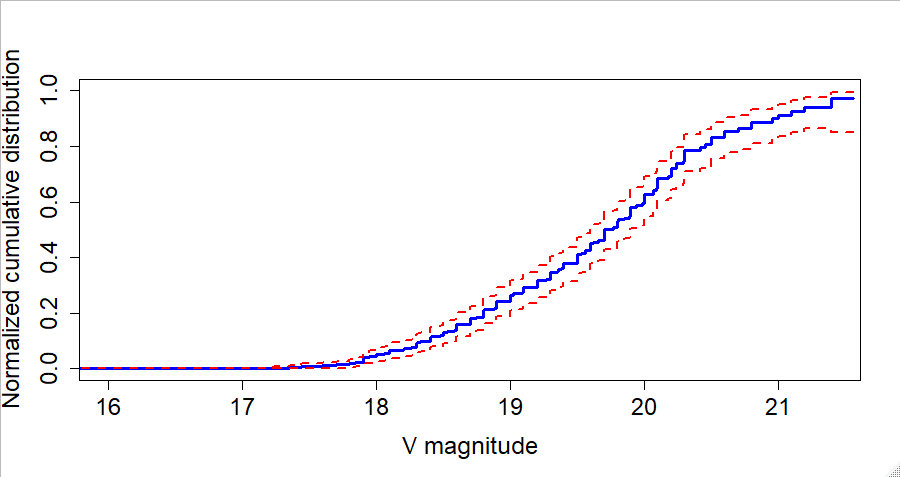} 
\caption{Distributions calculated from the survival function with 95\% confidence bounds in the optical V band.}
\label{fig:result_survivalfunc}
\end{figure}

We used the optical V magnitudes of Swift GRB observations, representing a sufficiently large homogeneous dataset sample, for a survival analysis in R programming language with 'survival' packages \cite{survival-package,survival-book}. This parameter is analogous to time in the survival analysis. Our dataset includes 437 GRBs V Magnitude measurements from the UVOT instrument, where 282 had an accurate observation and 155 were lower limit (censored) events. The difference between the total of the censored and accurate data and the number of gamma-ray bursts studied is due to the fact that for 15 GRBs, one of the parameters we used was missing or inaccurate, hence these 15 bursts were eliminated from further research. In Figure~\ref{fig:result_survival_compre}, we show the distribution of the exact measurements alongside the distribution of both the exact and censored measurements (all dataset). From the figures, we can clearly see that the two curves run together up to a magnitude of 17.5, after which the distributions diverge significantly due to the predicted limiting measurements. We, therefore, calculated the survival function from this magnitude. Then, using this survival function, we derived the traditional cumulative distribution function shown in Figure~\ref{fig:result_survivalfunc}. This figure presents the cumulative distribution of gamma bursts in magnitude V, along with the generally used statistical 95\% confidence interval. 

In order to better understand the results, we want to emphasise that the calculated Kaplan-Meier function is a mathematical approximation of the true survival function. This is important to remember when comparing it with the simple empirical cumulative distribution function (ECDF). The strength of the KM estimator lies in its ability to take into account the information inherent in the censored data, which is crucial for the survival analysis. Unlike the ECDF, which only takes into account fully observed data, the KM estimator corrects for the presence of censored observations, giving a more accurate picture of the true distribution of the observed data. This is particularly relevant in our study, as a non-negligible fraction of the magnitude V data provides only a lower bound and is, therefore, right-censored. By applying the survival analysis method, we can recover the true distribution function, taking into account all available information, not just the distribution that can be directly computed from successfully observed data. This distinction ensures that our survival analysis truly reflects the underlying patterns and relationships in the data.

\begin{figure}
\centering
\includegraphics[width=\linewidth]{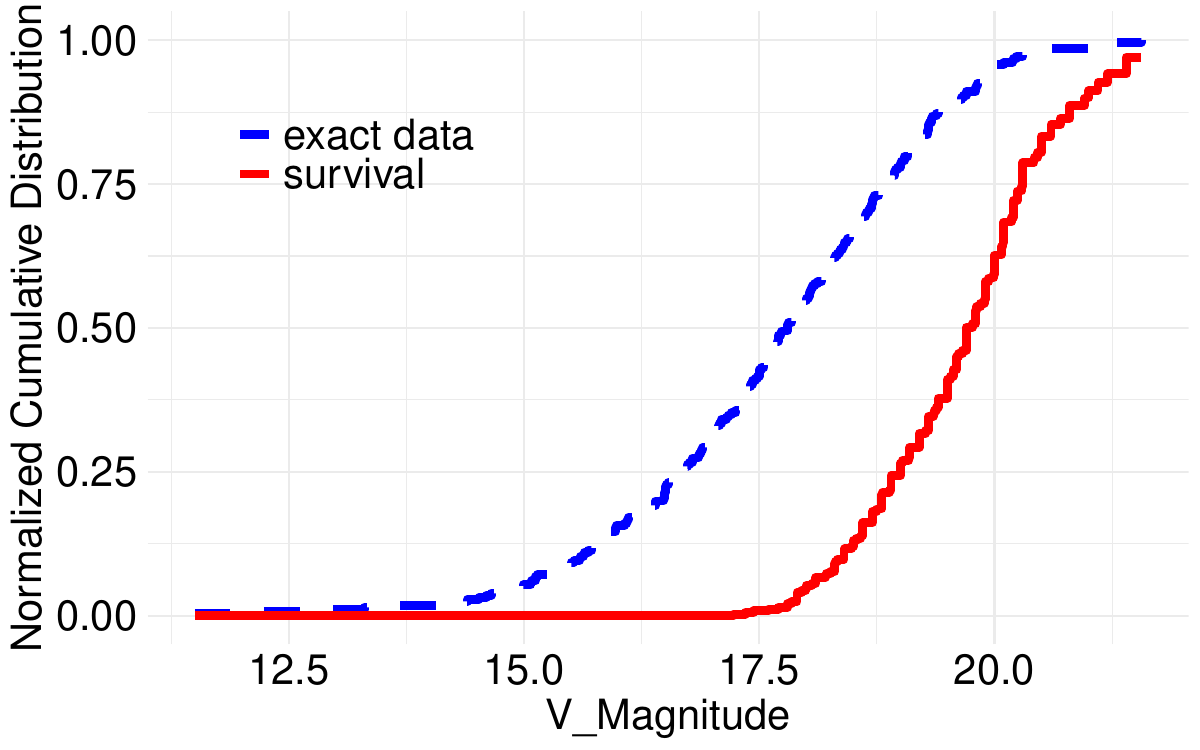} 
\caption{The figure shows how the survival function (red line) obtained using censored data differs from the cumulative distribution calculated from the exact data only. There is a significant difference between the two distributions. }
\label{fig:result_survival}
\end{figure}

We examined whether we can indeed obtain a significantly different distribution of outbreaks. Figure~\ref{fig:result_survival} clearly shows how different the two distributions are.
We used the Kolmogorov-Smirnov test \cite{Karson1968-ne} to compare the uncensored cumulative distribution with the survival function. The distance between the two distributions is 0.48, which is found to be a significant difference (significance level <2.2e-16 ($\approx8\sigma$)). That is, using the censored data, there is a significantly different statistical distribution of the brightness of GRBs in the V band.

\section{Discussion}
\subsection{Cox-regression}

The Cox regression was used to determine whether the properties of GRBs (SWIFT BAT's derived T90, Fluence, 1-sec Peak Photon Flux, Photon Index, XRT's derived 11 Hour Flux, gamma) depend on the visible magnitude distribution.
Table~\ref{tab:cox-regression} shows the coefficients of the selected physical parameters, as well as the probabilities of deviation from random chance.

\begin{table}
\centering
\begin{tabular}{|c|c|c|c|}
\hline 
Variable &$\mathrm{\beta}$ & HR $\mathrm{e^\beta}$ & Prob \\  
\hline 
$\log(BAT\ T90)$ & -0.20283 & 0.81642 & 0.392 \\ 
\hline 
$\log(BAT\ Fluence)$ & 0.34055  & 1.40572 &  0.312 \\ 
\hline 
$\log(BAT\ PeakFlux)$  & 0.04513  &  1.04616 & 0.890 \\ 
\hline 
BAT Photon index & 0.24081 &  1.27227 & 0.139 \\ 
\hline 
$\log(XRT\ 11hFlux)$ & 0.05693 & 1.05858  & 0.598 \\ 
\hline 
$\log(XRT\ Gamma)$ & -1.70346 & 0.18205 &0.235 \\ 
\hline 
\end{tabular} 
\caption{Results of the Cox regressions. The Prob column gives the probability, the dependence values are fully random. Seemingly, no remarkable correlation was obtained.}
\label{tab:cox-regression}
\end{table}

\subsection{Differences between $\gamma$-spectrum groups}

We compared the GRBs with simple power-law (PL) and cut-off power-law (CPL) spectral types. The defined p-value quantifies the probability of obtaining the observed data or more extreme results if the null hypothesis was true. It helps to determine whether the observed results are statistically significant or occurred by chance. A small p-value (typically less than 0.05) suggests strong evidence against the null hypothesis, while a large p-value indicates weak evidence against it. Applying this to our data, the number of GRBs with PL spectra is 359, and with CPL spectra 73, with the number of events ('censored data') in each group equal to 137 and 17, respectively. The p-value of 0.2 obtained indicates that there is no statistically significant difference between the CPL and PL spectra.

\section{Summary}

Survival analysis offers a robust statistical framework for examining the time until an event occurs, making it widely applicable across various fields, including medical research, engineering, and astrophysics. In the context of gamma-ray bursts (GRBs), survival analysis can be used to study the temporal characteristics of these explosive events, such as the duration of their different phases and the time intervals between various observed phenomena. By applying survival analysis techniques, researchers can gain deeper insights into the underlying mechanisms driving GRBs, and better understand the factors influencing their observed properties.

Previously, the optical luminosity distribution of GRBs was found to be significantly affected by several GRB properties, including duration, gamma-ray fluence and peak flux. This is explained by the fact that these factors play a crucial role in determining how bright the afterglow of a GRB will appear at the optical wavelength. The duration of the gamma-ray emission can affect the amount of energy deposited in the surrounding interstellar medium, while the gamma-ray fluence and peak flux are directly related to the total energy output and the intensity of the burst. These properties are indicative of the total energy budget of the GRB, which in turn affects the optical afterglow produced. However, in the present work, we have not found a significant correlation, so clarification of this should be the subject of further research.

We found 359 GRBs with power-laws pectra and 73 GRBs with cutoff power-law spectra in the Swift telescope database. This pattern is consistent with earlier research in the field, which found that PL spectra often explain a significantly larger percentage of GRB events than CPL spectra. PL spectra are more common than CPL spectra, which are generally linked to high-energy burst events or bursts with notable spectral evolution, according to studies such as those by \cite{2014ApJS..211...12G, 2018MNRAS.475..306R}. Our results support these patterns and further demonstrate that GRB spectral classification is consistent across various datasets and analytic techniques.

Interestingly, early X-ray flux and gamma photon index do not show a significant impact on the optical brightness distribution of GRBs. This suggests that the initial high-energy processes captured by X-ray observations may not directly correlate with the optical afterglow properties. Instead, the energetics of the jet launched from the central engine of the GRB likely play a more pivotal role. The interaction of this jet with the surrounding interstellar matter triggers the afterglow, with the gamma-ray properties reflecting the energy and dynamics of this interaction. Understanding the connection between these gamma-ray properties and the resulting optical brightness can provide valuable insights into the physics of GRB jets and their interactions with the environment.

\begin{acknowledgements}
The authors would like to thank the Hungarian TKP2021-NVA-16 programme for their support. Supported by the EKÖP-24-4-II-48 University Research Scholarship Program of the Ministry for Culture and Innovation from the source of the National Research, Development and Innovation Fund. The authors would like to thank L. Viktor Tóth for his help in the preparation of this article.
\end{acknowledgements}

\bibliographystyle{actapoly}
\bibliography{bib}

@ARTICLE{Kumar,
   author = {{Kumar}, P. and {Zhang}, B.},
    title = "{The physics of gamma-ray bursts \& relativistic jets}",
  journal = {\physrep},
archivePrefix = "arXiv",
   eprint = {1410.0679},
 primaryClass = "astro-ph.HE",
     year = 2015,
    month = feb,
   volume = 561,
    pages = {1-109},
      doi = {10.1016/j.physrep.2014.09.008},
   adsurl = {http://adsabs.harvard.edu/abs/2015PhR...561....1K}
}

@ARTICLE{meszaros_2006,
   author = {{M{\'e}sz{\'a}ros}, P.},
    title = "{Gamma-ray bursts}",
  journal = {Reports on Progress in Physics},
   eprint = {astro-ph/0605208},
     year = 2006,
    month = aug,
   volume = 69,
    pages = {2259-2321},
      doi = {10.1088/0034-4885/69/8/R01},
   adsurl = {http://adsabs.harvard.edu/abs/2006RPPh...69.2259M}
}

@ARTICLE{meszaros_fireball,
   author = {{Zhang}, B. and {M{\'e}sz{\'a}ros}, P.},
    title = "{An Analysis of Gamma-Ray Burst Spectral Break Models}",
  journal = {\apj},
   eprint = {astro-ph/0206158},
     year = 2002,
    month = dec,
   volume = 581,
    pages = {1236-1247},
      doi = {10.1086/344338},
   adsurl = {http://adsabs.harvard.edu/abs/2002ApJ...581.1236Z}
}

@ARTICLE{collapsar,
   author = {{MacFadyen}, A.~I. and {Woosley}, S.~E.},
    title = "{Collapsars: Gamma-Ray Bursts and Explosions in ``Failed Supernovae''}",
  journal = {\apj},
   eprint = {astro-ph/9810274},
     year = 1999,
    month = oct,
   volume = 524,
    pages = {262-289},
      doi = {10.1086/307790},
   adsurl = {http://adsabs.harvard.edu/abs/1999ApJ...524..262M}
}

@ARTICLE{ns_merging,
   author = {{Eichler}, D. and {Livio}, M. and {Piran}, T. and {Schramm}, D.~N.
	},
    title = "{Nucleosynthesis, neutrino bursts and gamma-rays from coalescing neutron stars}",
  journal = {\nat},
     year = 1989,
    month = jul,
   volume = 340,
    pages = {126-128},
      doi = {10.1038/340126a0},
   adsurl = {http://adsabs.harvard.edu/abs/1989Natur.340..126E}
}

@ARTICLE{veres_gw170814,
   author = {{Goldstein}, A. and {Veres}, P. and {Burns}, E. and {Briggs}, M.~S. and 
	{Hamburg}, R. and {Kocevski}, D. and {Wilson-Hodge}, C.~A. and 
	{Preece}, R.~D. and {Poolakkil}, S. and {Roberts}, O.~J. and 
	{Hui}, C.~M. and {Connaughton}, V. and {Racusin}, J. and {von Kienlin}, A. and 
	{Dal Canton}, T. and {Christensen}, N. and {Littenberg}, T. and 
	{Siellez}, K. and {Blackburn}, L. and {Broida}, J. and {Bissaldi}, E. and 
	{Cleveland}, W.~H. and {Gibby}, M.~H. and {Giles}, M.~M. and 
	{Kippen}, R.~M. and {McBreen}, S. and {McEnery}, J. and {Meegan}, C.~A. and 
	{Paciesas}, W.~S. and {Stanbro}, M.},
    title = "{An Ordinary Short Gamma-Ray Burst with Extraordinary Implications: Fermi-GBM Detection of GRB 170817A}",
  journal = {\apjl},
archivePrefix = "arXiv",
   eprint = {1710.05446},
 primaryClass = "astro-ph.HE",
     year = 2017,
    month = oct,
   volume = 848,
      eid = {L14},
    pages = {L14},
      doi = {10.3847/2041-8213/aa8f41},
   adsurl = {http://adsabs.harvard.edu/abs/2017ApJ...848L..14G}
}

@ARTICLE{hoi1998,
   author = {{Horv{\'a}th}, I.},
    title = "{A Third Class of Gamma-Ray Bursts?}",
  journal = {\apj},
   eprint = {astro-ph/9803077},
     year = 1998,
    month = dec,
   volume = 508,
    pages = {757-759},
      doi = {10.1086/306416},
   adsurl = {http://cdsads.u-strasbg.fr/abs/1998ApJ...508..757H}
}

@ARTICLE{hoi2006,
   author = {{Horv{\'a}th}, I. and {Bal{\'a}zs}, L.~G. and {Bagoly}, Z. and 
	{Ryde}, F. and {M{\'e}sz{\'a}ros}, A.},
    title = "{A new definition of the intermediate group of gamma-ray bursts}",
  journal = {\aap},
   eprint = {astro-ph/0509909},
     year = 2006,
    month = feb,
   volume = 447,
    pages = {23-30},
      doi = {10.1051/0004-6361:20041129},
   adsurl = {http://adsabs.harvard.edu/abs/2006A%26A...447...23H}
}

@ARTICLE{hoi2010,
       author = {{Horv{\'a}th}, I. and {Bagoly}, Z. and {Bal{\'a}zs}, L.~G. and {de
        Ugarte Postigo}, A. and {Veres}, P. and {M{\'e}sz{\'a}ros}, A.},
        title = "{Detailed Classification of Swift 's Gamma-ray Bursts}",
      journal = {\apj},
         year = 2010,
        month = Apr,
       volume = {713},
        pages = {552-557},
          doi = {10.1088/0004-637X/713/1/552},
archivePrefix = {arXiv},
       eprint = {1003.0632},
 primaryClass = {astro-ph.HE},
       adsurl = {https://ui.adsabs.harvard.edu/#abs/2010ApJ...713..552H}
}

@ARTICLE{Xiongwei2018,
   author = {{Bi}, X. and {Mao}, J. and {Liu}, C. and {Bai}, J.-M.},
    title = "{Statistical Study of the Swift X-Ray Flash and X-Ray Rich Gamma-Ray Bursts}",
  journal = {\apj},
archivePrefix = "arXiv",
   eprint = {1810.03079},
 primaryClass = "astro-ph.HE",
     year = 2018,
    month = oct,
   volume = 866,
      eid = {97},
    pages = {97},
      doi = {10.3847/1538-4357/aadcf8},
   adsurl = {http://adsabs.harvard.edu/abs/2018ApJ...866...97B}
}

@article{pinter_2018,
title={Resolving the structure of the Galactic foreground using Herschel measurements and the Kriging technique}, 
    volume={12},
    number={S333}, 
   journal={Proceedings of the International Astronomical Union}, 
 publisher={Cambridge University Press}, 
    author={Pinter, S. and Bagoly, Z. and Balázs, L. G. and Horvath, I. and Racz, I. I. and Zahorecz, S. and Tóth, L. V.}, 
      year={2017}, 
     pages={168–169},
      doi = {10.1017/S1743921317011097},
   adsurl = {https://ui.adsabs.harvard.edu/abs/2018IAUS..333..168P}
}

@ARTICLE{swift_bat_2005,
   author = {{Barthelmy}, S.~D. and {Barbier}, L.~M. and {Cummings}, J.~R. and 
	{Fenimore}, E.~E. and {Gehrels}, N. and {Hullinger}, D. and 
	{Krimm}, H.~A. and {Markwardt}, C.~B. and {Palmer}, D.~M. and 
	{Parsons}, A. and {Sato}, G. and {Suzuki}, M. and {Takahashi}, T. and 
	{Tashiro}, M. and {Tueller}, J.},
    title = "{The Burst Alert Telescope (BAT) on the SWIFT Midex Mission}",
  journal = {Space Science Reviews},
   eprint = {astro-ph/0507410},
     year = 2005,
    month = oct,
   volume = 120,
    pages = {143-164},
      doi = {10.1007/s11214-005-5096-3},
   adsurl = {http://adsabs.harvard.edu/abs/2005SSRv..120..143B}
}

@ARTICLE{uvot_2005,
   author = {{Roming}, P.~W.~A. and {Kennedy}, T.~E. and {Mason}, K.~O. and 
	{Nousek}, J.~A. and {Ahr}, L. and {Bingham}, R.~E. and {Broos}, P.~S. and 
	{Carter}, M.~J. and {Hancock}, B.~K. and {Huckle}, H.~E. and 
	{Hunsberger}, S.~D. and {Kawakami}, H. and {Killough}, R. and 
	{Koch}, T.~S. and {McLelland}, M.~K. and {Smith}, K. and {Smith}, P.~J. and 
	{Soto}, J.~C. and {Boyd}, P.~T. and {Breeveld}, A.~A. and {Holland}, S.~T. and 
	{Ivanushkina}, M. and {Pryzby}, M.~S. and {Still}, M.~D. and 
	{Stock}, J.},
    title = "{The Swift Ultra-Violet/Optical Telescope}",
  journal = {Space Science Reviews},
   eprint = {astro-ph/0507413},
     year = 2005,
    month = oct,
   volume = 120,
    pages = {95-142},
      doi = {10.1007/s11214-005-5095-4},
   adsurl = {http://adsabs.harvard.edu/abs/2005SSRv..120...95R}
}

@ARTICLE{xrt_2005,
   author = {{Burrows}, D.~N. and {Hill}, J.~E. and {Nousek}, J.~A. and {Kennea}, J.~A. and 
	{Wells}, A. and {Osborne}, J.~P. and {Abbey}, A.~F. and {Beardmore}, A. and 
	{Mukerjee}, K. and {Short}, A.~D.~T. and {Chincarini}, G. and 
	{Campana}, S. and {Citterio}, O. and {Moretti}, A. and {Pagani}, C. and 
	{Tagliaferri}, G. and {Giommi}, P. and {Capalbi}, M. and {Tamburelli}, F. and 
	{Angelini}, L. and {Cusumano}, G. and {Br{\"a}uninger}, H.~W. and 
	{Burkert}, W. and {Hartner}, G.~D.},
    title = "{The Swift X-Ray Telescope}",
  journal = {Space Science Reviews},
   eprint = {astro-ph/0508071},
     year = 2005,
    month = oct,
   volume = 120,
    pages = {165-195},
      doi = {10.1007/s11214-005-5097-2},
   adsurl = {http://adsabs.harvard.edu/abs/2005SSRv..120..165B}
}

@ARTICLE{great_wall_1,
       author = {{Horv{\'a}th}, Istv{\'a}n and {Hakkila}, Jon and {Bagoly}, Zsolt},
        title = "{Possible structure in the GRB sky distribution at redshift two}",
      journal = {\aap},
         year = 2014,
        month = Jan,
       volume = {561},
          eid = {L12},
        pages = {L12},
          doi = {10.1051/0004-6361/201323020},
archivePrefix = {arXiv},
       eprint = {1401.0533},
 primaryClass = {astro-ph.CO},
       adsurl = {https://ui.adsabs.harvard.edu/#abs/2014A&A...561L..12H}
}

@ARTICLE{great_wall_2,
       author = {{Horv{\'a}th}, Istv{\'a}n and {Bagoly}, Zsolt and {Hakkila}, Jon and
        {T{\'o}th}, L.~V.},
        title = "{New data support the existence of the Hercules-Corona Borealis Great
        Wall}",
      journal = {\aap},
         year = 2015,
        month = Dec,
       volume = {584},
          eid = {A48},
        pages = {A48},
          doi = {10.1051/0004-6361/201424829},
 primaryClass = {astro-ph.HE},
       adsurl = {https://ui.adsabs.harvard.edu/#abs/2015A&A...584A..48H}
}

@ARTICLE{ring_1,
       author = {{Bal{\'a}zs}, L.~G. and {Bagoly}, Z. and {Hakkila}, J.~E. and
        {Horv{\'a}th}, I. and {K{\'o}bori}, J. and {R{\'a}cz}, I.~I. and
        {T{\'o}th}, L.~V.},
        title = "{A giant ring-like structure at $0.78 < z < 0.86$ displayed by GRBs}",
      journal = {\mnras},
         year = 2015,
        month = Sep,
       volume = {452},
        pages = {2236-2246},
          doi = {10.1093/mnras/stv1421},
 primaryClass = {astro-ph.CO},
       adsurl = {https://ui.adsabs.harvard.edu/#abs/2015MNRAS.452.2236B}
}

@ARTICLE{ring_2,
       author = {{Bal{\'a}zs}, Lajos G. and {Rejt{\'{o}}}, L{\'\i}dia and {Tusn{\'a}dy},
        G{\'a}bor},
        title = "{Some statistical remarks on the giant GRB ring}",
      journal = {\mnras},
         year = 2018,
        month = Jan,
       volume = {473},
        pages = {3169-3179},
          doi = {10.1093/mnras/stx2550},
 primaryClass = {astro-ph.CO},
       adsurl = {https://ui.adsabs.harvard.edu/#abs/2018MNRAS.473.3169B}
}

@ARTICLE{farthest_grb_spec,
       author = {{Tanvir}, N.~R. and {Fox}, D.~B. and {Levan}, A.~J. and {Berger}, E. and
        {Wiersema}, K. and {Fynbo}, J.~P.~U. and {Cucchiara}, A. and
        {Kr{\"u}hler}, T. and {Gehrels}, N. and {Bloom}, J.~S. and
        {Greiner}, J. and {Evans}, P.~A. and {Rol}, E. and {Olivares},
        F. and {Hjorth}, J. and {Jakobsson}, P. and {Farihi}, J. and
        {Willingale}, R. and {Starling}, R.~L.~C. and {Cenko}, S.~B. and
        {Perley}, D. and {Maund}, J.~R. and {Duke}, J. and {Wijers},
        R.~A.~M.~J. and {Adamson}, A.~J. and {Allan}, A. and {Bremer},
        M.~N. and {Burrows}, D.~N. and {Castro-Tirado}, A.~J. and
        {Cavanagh}, B. and {de Ugarte Postigo}, A. and {Dopita}, M.~A.
        and {Fatkhullin}, T.~A. and {Fruchter}, A.~S. and {Foley}, R.~J.
        and {Gorosabel}, J. and {Kennea}, J. and {Kerr}, T. and {Klose},
        S. and {Krimm}, H.~A. and {Komarova}, V.~N. and {Kulkarni},
        S.~R. and {Moskvitin}, A.~S. and {Mundell}, C.~G. and {Naylor},
        T. and {Page}, K. and {Penprase}, B.~E. and {Perri}, M. and
        {Podsiadlowski}, P. and {Roth}, K. and {Rutledge}, R.~E. and
        {Sakamoto}, T. and {Schady}, P. and {Schmidt}, B.~P. and
        {Soderberg}, A.~M. and {Sollerman}, J. and {Stephens}, A.~W. and
        {Stratta}, G. and {Ukwatta}, T.~N. and {Watson}, D. and
        {Westra}, E. and {Wold}, T. and {Wolf}, C.},
        title = "{A {\ensuremath{\gamma}}-ray burst at a redshift of z\textasciitilde8.2}",
      journal = {\nat},
         year = 2009,
        month = Oct,
       volume = {461},
        pages = {1254-1257},
          doi = {10.1038/nature08459},
archivePrefix = {arXiv},
       eprint = {0906.1577},
 primaryClass = {astro-ph.CO},
       adsurl = {https://ui.adsabs.harvard.edu/#abs/2009Natur.461.1254T}
}

@ARTICLE{swom,
   author = {{Zhao}, D. and {Cordier}, B. and {Sizun}, P. and {Wu}, B. and 
	{Dong}, Y. and {Schanne}, S. and {Song}, L. and {Liu}, J.},
    title = "{Influence of the Earth on the background and the sensitivity of the GRM and ECLAIRs instruments aboard the Chinese-French mission SVOM}",
  journal = {Experimental Astronomy},
archivePrefix = "arXiv",
   eprint = {1208.2493},
 primaryClass = "astro-ph.IM",
     year = 2012,
    month = nov,
   volume = 34,
    pages = {705-728},
      doi = {10.1007/s10686-012-9313-2},
   adsurl = {http://adsabs.harvard.edu/abs/2012ExA....34..705Z}
}

@ARTICLE{theseus01,
   author = {{Amati}, L. and {O'Brien}, P. and {G{\"o}tz}, D. and {Bozzo}, E. and 
	{Tenzer}, C. and {Frontera}, F. and {Ghirlanda}, G. and {Labanti}, C. and 
	{Osborne}, J.~P. and {Stratta}, G. and {Tanvir}, N. and {Willingale}, R. and 
	{Attina}, P. and {Campana}, R. and {Castro-Tirado}, A.~J. and 
	{Contini}, C. and {Fuschino}, F. and {Gomboc}, A. and {Hudec}, R. and 
	{Orleanski}, P. and {Renotte}, E. and {Rodic}, T. and {Bagoly}, Z. and 
	{Blain}, A. and {Callanan}, P. and {Covino}, S. and {Ferrara}, A. and 
	{Le Floch}, E. and {Marisaldi}, M. and {Mereghetti}, S. and 
	{Rosati}, P. and {Vacchi}, A. and {D'Avanzo}, P. and {Giommi}, P. and 
	{Piranomonte}, S. and {Piro}, L. and {Reglero}, V. and {Rossi}, A. and 
	{Santangelo}, A. and {Salvaterra}, R. and {Tagliaferri}, G. and 
	{Vergani}, S. and {Vinciguerra}, S. and {Briggs}, M. and {Campolongo}, E. and 
	{Ciolfi}, R. and {Connaughton}, V. and {Cordier}, B. and {Morelli}, B. and 
	{Orlandini}, M. and {Adami}, C. and {Argan}, A. and {Atteia}, J.-L. and 
	{Auricchio}, N. and {Balazs}, L. and {Baldazzi}, G. and {Basa}, S. and 
	{Basak}, R. and {Bellutti}, P. and {Bernardini}, M.~G. and {Bertuccio}, G. and 
	{Braga}, J. and {Branchesi}, M. and {Brandt}, S. and {Brocato}, E. and 
	{Budtz-Jorgensen}, C. and {Bulgarelli}, A. and {Burderi}, L. and 
	{Camp}, J. and {Capozziello}, S. and {Caruana}, J. and {Casella}, P. and 
	{Cenko}, B. and {Chardonnet}, P. and {Ciardi}, B. and {Colafrancesco}, S. and 
	{Dainotti}, M.~G. and {D'Elia}, V. and {De Martino}, D. and 
	{De Pasquale}, M. and {Del Monte}, E. and {Della Valle}, M. and 
	{Drago}, A. and {Evangelista}, Y. and {Feroci}, M. and {Finelli}, F. and 
	{Fiorini}, M. and {Fynbo}, J. and {Gal-Yam}, A. and {Gendre}, B. and 
	{Ghisellini}, G. and {Grado}, A. and {Guidorzi}, C. and {Hafizi}, M. and 
	{Hanlon}, L. and {Hjorth}, J. and {Izzo}, L. and {Kiss}, L. and 
	{Kumar}, P. and {Kuvvetli}, I. and {Lavagna}, M. and {Li}, T. and 
	{Longo}, F. and {Lyutikov}, M. and {Maio}, U. and {Maiorano}, E. and 
	{Malcovati}, P. and {Malesani}, D. and {Margutti}, R. and {Martin-Carrillo}, A. and 
	{Masetti}, N. and {McBreen}, S. and {Mignani}, R. and {Morgante}, G. and 
	{Mundell}, C. and {Nargaard-Nielsen}, H.~U. and {Nicastro}, L. and 
	{Palazzi}, E. and {Paltani}, S. and {Panessa}, F. and {Pareschi}, G. and 
	{Pe'er}, A. and {Penacchioni}, A.~V. and {Pian}, E. and {Piedipalumbo}, E. and 
	{Piran}, T. and {Rauw}, G. and {Razzano}, M. and {Read}, A. and 
	{Rezzolla}, L. and {Romano}, P. and {Ruffini}, R. and {Savaglio}, S. and 
	{Sguera}, V. and {Schady}, P. and {Skidmore}, W. and {Song}, L. and 
	{Stanway}, E. and {Starling}, R. and {Topinka}, M. and {Troja}, E. and 
	{van Putten}, M. and {Vanzella}, E. and {Vercellone}, S. and 
	{Wilson-Hodge}, C. and {Yonetoku}, D. and {Zampa}, G. and {Zampa}, N. and 
	{Zhang}, B. and {Zhang}, B.~B. and {Zhang}, S. and {Zhang}, S.-N. and 
	{Antonelli}, A. and {Bianco}, F. and {Boci}, S. and {Boer}, M. and 
	{Botticella}, M.~T. and {Boulade}, O. and {Butler}, C. and {Campana}, S. and 
	{Capitanio}, F. and {Celotti}, A. and {Chen}, Y. and {Colpi}, M. and 
	{Comastri}, A. and {Cuby}, J.-G. and {Dadina}, M. and {De Luca}, A. and 
	{Dong}, Y.-W. and {Ettori}, S. and {Gandhi}, P. and {Geza}, E. and 
	{Greiner}, J. and {Guiriec}, S. and {Harms}, J. and {Hernanz}, M. and 
	{Hornstrup}, A. and {Hutchinson}, I. and {Israel}, G. and {Jonker}, P. and 
	{Kaneko}, Y. and {Kawai}, N. and {Wiersema}, K. and {Korpela}, S. and 
	{Lebrun}, V. and {Lu}, F. and {MacFadyen}, A. and {Malaguti}, G. and 
	{Maraschi}, L. and {Melandri}, A. and {Modjaz}, M. and {Morris}, D. and 
	{Omodei}, N. and {Paizis}, A. and {P{\'a}ta}, P. and {Petrosian}, V. and 
	{Rachevski}, A. and {Rhoads}, J. and {Ryde}, F. and {Sabau-Graziati}, L. and 
	{Shigehiro}, N. and {Sims}, M. and {Soomin}, J. and {Sz{\'e}csi}, D. and 
	{Urata}, Y. and {Uslenghi}, M. and {Valenziano}, L. and {Vianello}, G. and 
	{Vojtech}, S. and {Watson}, D. and {Zicha}, J.},
    title = "{The THESEUS space mission concept: science case, design and expected performances}",
  journal = {Advances in Space Research},
archivePrefix = "arXiv",
   eprint = {1710.04638},
 primaryClass = "astro-ph.IM",
     year = 2018,
    month = jul,
   volume = 62,
    pages = {191-244},
      doi = {10.1016/j.asr.2018.03.010},
   adsurl = {http://adsabs.harvard.edu/abs/2018AdSpR..62..191A}
}

@book{zhang_2018, 
    place = {Cambridge}, 
    title = {The Physics of Gamma-Ray Bursts}, 
    DOI = {10.1017/9781139226530}, 
    publisher = {Cambridge University Press}, 
    author = {Zhang, Bing}, 
    year = {2018}
}

@ARTICLE{2008MNRAS.391.1741V,
       author = {{Vavrek}, R. and {Bal{\'a}zs}, L.~G. and {M{\'e}sz{\'a}ros}, A. and {Horv{\'a}th}, I. and {Bagoly}, Z.},
        title = "{Testing the randomness in the sky-distribution of gamma-ray bursts}",
      journal = {\mnras},
         year = 2008,
        month = dec,
       volume = {391},
       number = {4},
        pages = {1741-1748},
          doi = {10.1111/j.1365-2966.2008.13635.x},
archivePrefix = {arXiv},
       eprint = {0806.3932},
 primaryClass = {astro-ph},
       adsurl = {https://ui.adsabs.harvard.edu/abs/2008MNRAS.391.1741V}
}

@ARTICLE{1999A&AS..138..417B,
       author = {{Bal{\'a}zs}, L.~G. and {M{\'e}sz{\'a}ros}, A. and {Horv{\'a}th}, I. and {Vavrek}, R.},
        title = "{An intrinsic anisotropy in the angular distribution of gamma-ray bursts}",
      journal = {\aaps},
         year = 1999,
        month = sep,
       volume = {138},
        pages = {417-418},
          doi = {10.1051/aas:1999290},
archivePrefix = {arXiv},
       eprint = {astro-ph/9909379},
 primaryClass = {astro-ph},
       adsurl = {https://ui.adsabs.harvard.edu/abs/1999A&AS..138..417B}
}

@ARTICLE{2019MNRAS.486.4823T,
       author = {{T{\'o}th}, B.~G. and {R{\'a}cz}, I.~I. and {Horv{\'a}th}, I.},
        title = "{Gaussian-mixture-model-based cluster analysis of gamma-ray bursts in the BATSE catalog}",
      journal = {\mnras},
         year = 2019,
        month = jul,
       volume = {486},
       number = {4},
        pages = {4823-4828},
          doi = {10.1093/mnras/stz1188},
archivePrefix = {arXiv},
       eprint = {1906.01362},
 primaryClass = {astro-ph.HE},
       adsurl = {https://ui.adsabs.harvard.edu/abs/2019MNRAS.486.4823T}
}

@ARTICLE{2020MNRAS.498.2544H,
       author = {{Horvath}, I. and {Sz{\'e}csi}, D. and {Hakkila}, J. and {Szab{\'o}}, {\'A}. and {Racz}, I.~I. and {T{\'o}th}, L.~V. and {Pinter}, S. and {Bagoly}, Z.},
        title = "{The clustering of gamma-ray bursts in the Hercules-Corona Borealis Great Wall: the largest structure in the Universe?}",
      journal = {\mnras},
         year = 2020,
        month = oct,
       volume = {498},
       number = {2},
        pages = {2544-2553},
          doi = {10.1093/mnras/staa2460},
archivePrefix = {arXiv},
       eprint = {2008.03679},
 primaryClass = {astro-ph.HE},
       adsurl = {https://ui.adsabs.harvard.edu/abs/2020MNRAS.498.2544H}
}

@ARTICLE{2019Ap&SS.364..105H,
       author = {{Horv{\'a}th}, I. and {Hakkila}, J. and {Bagoly}, Z. and {T{\'o}th}, L.~V. and {R{\'a}cz}, I.~I. and {Pint{\'e}r}, S. and {T{\'o}th}, B.~G.},
        title = "{Multidimensional analysis of Fermi GBM gamma-ray bursts}",
      journal = {\apss},
         year = 2019,
        month = jun,
       volume = {364},
       number = {6},
          eid = {105},
        pages = {105},
          doi = {10.1007/s10509-019-3585-1},
archivePrefix = {arXiv},
       eprint = {1905.12998},
 primaryClass = {astro-ph.HE},
       adsurl = {https://ui.adsabs.harvard.edu/abs/2019Ap&SS.364..105H}
}

@ARTICLE{2018ApJ...855..101H,
       author = {{Hakkila}, Jon and {Horv{\'a}th}, Istv{\'a}n and {Hofesmann}, Eric and {Lesage}, Stephen},
        title = "{Properties of Short Gamma-ray Burst Pulses from a BATSE TTE GRB Pulse Catalog}",
      journal = {\apj},
         year = 2018,
        month = mar,
       volume = {855},
       number = {2},
          eid = {101},
        pages = {101},
          doi = {10.3847/1538-4357/aaac2b},
archivePrefix = {arXiv},
       eprint = {1710.08957},
 primaryClass = {astro-ph.HE},
       adsurl = {https://ui.adsabs.harvard.edu/abs/2018ApJ...855..101H}
}

@ARTICLE{2010A&A...510A.105P,
       author = {{P{\'e}rez-Ram{\'\i}rez}, D. and {de Ugarte Postigo}, A. and {Gorosabel}, J. and {Aloy}, M.~A. and {J{\'o}hannesson}, G. and {Guerrero}, M.~A. and {Osborne}, J.~P. and {Page}, K.~L. and {Warwick}, R.~S. and {Horv{\'a}th}, I. and {Veres}, P. and {Jel{\'\i}nek}, M. and {Kub{\'a}nek}, P. and {Guziy}, S. and {Bremer}, M. and {Winters}, J.~M. and {Riva}, A. and {Castro-Tirado}, A.~J.},
        title = "{Detection of the high z GRB 080913 and its implications on progenitors and energy extraction mechanisms}",
      journal = {\aap},
         year = 2010,
        month = feb,
       volume = {510},
          eid = {A105},
        pages = {A105},
          doi = {10.1051/0004-6361/200811151},
archivePrefix = {arXiv},
       eprint = {0810.2107},
 primaryClass = {astro-ph},
       adsurl = {https://ui.adsabs.harvard.edu/abs/2010A&A...510A.105P}
}

@ARTICLE{balazs1998,
       author = {{Balazs}, L.~G. and {Meszaros}, A. and {Horvath}, I.},
        title = "{Anisotropy of the sky distribution of gamma-ray bursts}",
      journal = {\aap},
         year = 1998,
        month = nov,
       volume = {339},
        pages = {1-6},
          doi = {10.48550/arXiv.astro-ph/9807006},
archivePrefix = {arXiv},
       eprint = {astro-ph/9807006},
 primaryClass = {astro-ph},
       adsurl = {https://ui.adsabs.harvard.edu/abs/1998A&A...339....1B}
}

@ARTICLE{2000ApJ...539...98M,
       author = {{M{\'e}sz{\'a}ros}, Attila and {Bagoly}, Zsolt and {Horv{\'a}th}, Istv{\'a}n and {Bal{\'a}zs}, Lajos G. and {Vavrek}, Roland},
        title = "{A Remarkable Angular Distribution of the Intermediate Subclass of Gamma-Ray Bursts}",
      journal = {\apj},
         year = 2000,
        month = aug,
       volume = {539},
       number = {1},
        pages = {98-101},
          doi = {10.1086/309193},
archivePrefix = {arXiv},
       eprint = {astro-ph/0003284},
 primaryClass = {astro-ph},
       adsurl = {https://ui.adsabs.harvard.edu/abs/2000ApJ...539...98M}
}

@ARTICLE{2000A&A...354....1M,
       author = {{M{\'e}sz{\'a}ros}, A. and {Bagoly}, Z. and {Vavrek}, R.},
        title = "{On the existence of the intrinsic anisotropies in the angular distributions of gamma-ray bursts}",
      journal = {\aap},
         year = 2000,
        month = feb,
       volume = {354},
        pages = {1-6},
          doi = {10.48550/arXiv.astro-ph/9912037},
archivePrefix = {arXiv},
       eprint = {astro-ph/9912037},
 primaryClass = {astro-ph},
       adsurl = {https://ui.adsabs.harvard.edu/abs/2000A&A...354....1M}
}

@ARTICLE{2022Univ....8..221H,
       author = {{Horvath}, Istvan and {Racz}, Istvan I. and {Bagoly}, Zsolt and {Bal{\'a}zs}, Lajos G. and {Pinter}, Sandor},
        title = "{Does the GRB Duration Depend on Redshift?}",
      journal = {Universe},
         year = 2022,
        month = mar,
       volume = {8},
       number = {4},
        pages = {221},
          doi = {10.3390/universe8040221},
archivePrefix = {arXiv},
       eprint = {2206.13621},
 primaryClass = {astro-ph.HE},
       adsurl = {https://ui.adsabs.harvard.edu/abs/2022Univ....8..221H}
}

@ARTICLE{hor04,
       author = {{Horv{\'a}th}, I. and {M{\'e}sz{\'a}ros}, A. and {Bal{\'a}zs}, L.~G. and {Bagoly}, Z.},
        title = "{Where is the Third Subgroup of Gamma-Ray Bursts?}",
      journal = {Baltic Astronomy},
         year = 2004,
        month = jan,
       volume = {13},
        pages = {217-220},
          doi = {10.48550/arXiv.astro-ph/0507688},
archivePrefix = {arXiv},
       eprint = {astro-ph/0507688},
 primaryClass = {astro-ph},
       adsurl = {https://ui.adsabs.harvard.edu/abs/2004BaltA..13..217H}
}

@ARTICLE{2018Ap&SS.363...53H,
       author = {{Horv{\'a}th}, I. and {T{\'o}th}, B.~G. and {Hakkila}, J. and {T{\'o}th}, L.~V. and {Bal{\'a}zs}, L.~G. and {R{\'a}cz}, I.~I. and {Pint{\'e}r}, S. and {Bagoly}, Z.},
        title = "{Classifying GRB 170817A/GW170817 in a Fermi duration-hardness plane}",
      journal = {\apss},
         year = 2018,
        month = mar,
       volume = {363},
       number = {3},
          eid = {53},
        pages = {53},
          doi = {10.1007/s10509-018-3274-5},
archivePrefix = {arXiv},
       eprint = {1710.11509},
 primaryClass = {astro-ph.CO},
       adsurl = {https://ui.adsabs.harvard.edu/abs/2018Ap&SS.363...53H}
}

@ARTICLE{2016A&A...593L..10B,
       author = {{Bagoly}, Zsolt and {Sz{\'e}csi}, Dorottya and {Bal{\'a}zs}, Lajos G. and {Csabai}, Istv{\'a}n and {Horv{\'a}th}, Istv{\'a}n and {Dobos}, L{\'a}szl{\'o} and {Lichtenberger}, J{\'a}nos and {T{\'o}th}, L. Viktor},
        title = "{Searching for electromagnetic counterpart of LIGO gravitational waves in the Fermi GBM data with ADWO}",
      journal = {\aap},
         year = 2016,
        month = sep,
       volume = {593},
          eid = {L10},
        pages = {L10},
          doi = {10.1051/0004-6361/201628569},
archivePrefix = {arXiv},
       eprint = {1603.06611},
 primaryClass = {astro-ph.HE},
       adsurl = {https://ui.adsabs.harvard.edu/abs/2016A&A...593L..10B}
}

@ARTICLE{2017CoSka..47...76B,
       author = {{Bagoly}, Z. and {Sz{\'e}csi}, D. and {Bal{\'a}zs}, L.~G. and {Csabai}, I. and {Dobos}, L. and {Horv{\'a}th}, I. and {Lichtenberger}, J. and {T{\'o}th}, L.~V.},
        title = "{Fermi GBM transient searches with ADWO}",
      journal = {Contributions of the Astronomical Observatory Skalnate Pleso},
         year = 2017,
        month = jul,
       volume = {47},
       number = {2},
        pages = {76-83},
       adsurl = {https://ui.adsabs.harvard.edu/abs/2017CoSka..47...76B}
}

@ARTICLE{2019AN....340..681B,
       author = {{Bagoly}, Zsolt and {Bal{\'a}zs}, Lajos G. and {Galg{\'o}czi}, G{\'a}bor and {Ohno}, Masanori and {P{\'a}l}, Andr{\'a}s. and {{\v{R}}{\'\i}pa}, Jakub and {T{\'o}th}, L. Viktor and {Werner}, Norbert},
        title = "{Transient detection capabilities of small satellite gamma-ray detectors}",
      journal = {Astronomische Nachrichten},
         year = 2019,
        month = aug,
       volume = {340},
       number = {7},
        pages = {681-689},
          doi = {10.1002/asna.201913675},
       adsurl = {https://ui.adsabs.harvard.edu/abs/2019AN....340..681B}
}

@ARTICLE{2020MNRAS.494.4343K,
       author = {{K{\'o}bori}, J{\'o}zsef and {Bagoly}, Zsolt and {Bal{\'a}zs}, Lajos G.},
        title = "{Kilonova rates from spherical and axisymmetrical models}",
      journal = {\mnras},
         year = 2020,
        month = may,
       volume = {494},
       number = {3},
        pages = {4343-4348},
          doi = {10.1093/mnras/staa1034},
archivePrefix = {arXiv},
       eprint = {2010.08041},
 primaryClass = {astro-ph.HE},
       adsurl = {https://ui.adsabs.harvard.edu/abs/2020MNRAS.494.4343K}
}

@ARTICLE{2021ExA....52..219T,
       author = {{Tanvir}, N.~R. and {Le Floc'h}, E. and {Christensen}, L. and {Caruana}, J. and {Salvaterra}, R. and {Ghirlanda}, G. and {Ciardi}, B. and {Maio}, U. and {D'Odorico}, V. and {Piedipalumbo}, E. and {Campana}, S. and {Noterdaeme}, P. and {Graziani}, L. and {Amati}, L. and {Bagoly}, Z. and {Bal{\'a}zs}, L.~G. and {Basa}, S. and {Behar}, E. and {De Cia}, A. and {Della Valle}, M. and {De Pasquale}, M. and {Frontera}, F. and {Gomboc}, A. and {G{\"o}tz}, D. and {Horvath}, I. and {Hudec}, R. and {Mereghetti}, S. and {O'Brien}, P.~T. and {Osborne}, J.~P. and {Paltani}, S. and {Rosati}, P. and {Sergijenko}, O. and {Stanway}, E.~R. and {Sz{\'e}csi}, D. and {Toth}, L.~V. and {Urata}, Y. and {Vergani}, S. and {Zane}, S.},
        title = "{Exploration of the high-redshift universe enabled by THESEUS}",
      journal = {Experimental Astronomy},
         year = 2021,
        month = dec,
       volume = {52},
       number = {3},
        pages = {219-244},
          doi = {10.1007/s10686-021-09778-w},
archivePrefix = {arXiv},
       eprint = {2104.09532},
 primaryClass = {astro-ph.IM},
       adsurl = {https://ui.adsabs.harvard.edu/abs/2021ExA....52..219T}
}

@ARTICLE{2019PASJ...71...10T,
       author = {{Toth}, L. Viktor and {Doi}, Yasuo and {Zahorecz}, Sarolta and {Pinter}, Sandor and {Racz}, Istvan I. and {Bagoly}, Zsolt and {Balazs}, Lajos G. and {Horvath}, Istvan and {Kiss}, Csaba and {Kov{\'a}cs}, T{\'\i}mea and {Onishi}, Toshikazu},
        title = "{Galactic foreground of gamma-ray bursts from AKARI Far-Infrared Surveyor}",
      journal = {\pasj},
         year = 2019,
        month = jan,
       volume = {71},
       number = {1},
          eid = {10},
        pages = {10},
          doi = {10.1093/pasj/psy123},
       adsurl = {https://ui.adsabs.harvard.edu/abs/2019PASJ...71...10T}
}

@ARTICLE{2010ApJ...725.1955V,
       author = {{Veres}, P. and {Bagoly}, Z. and {Horv{\'a}th}, I. and {M{\'e}sz{\'a}ros}, A. and {Bal{\'a}zs}, L.~G.},
        title = "{A Distinct Peak-flux Distribution of the Third Class of Gamma-ray Bursts: A Possible Signature of X-ray Flashes?}",
      journal = {\apj},
         year = 2010,
        month = dec,
       volume = {725},
       number = {2},
        pages = {1955-1964},
          doi = {10.1088/0004-637X/725/2/1955},
       adsurl = {https://ui.adsabs.harvard.edu/abs/2010ApJ...725.1955V}
}

@ARTICLE{2009A&A...493...51B,
       author = {{Bagoly}, Z. and {Borgonovo}, L. and {M{\'e}sz{\'a}ros}, A. and {Bal{\'a}zs}, L.~G. and {Horv{\'a}th}, I.},
        title = "{Factor analysis of the long gamma-ray bursts}",
      journal = {\aap},
         year = 2009,
        month = jan,
       volume = {493},
       number = {1},
        pages = {51-54},
          doi = {10.1051/0004-6361:20078635},
       adsurl = {https://ui.adsabs.harvard.edu/abs/2009A&A...493...51B}
}

@ARTICLE{2018MNRAS.475..306R,
       author = {{R{\'a}cz}, Istv{\'a}n I. and {Bal{\'a}zs}, Lajos G. and {Horvath}, Istvan and {T{\'o}th}, L. Viktor and {Bagoly}, Zsolt},
        title = "{Statistical properties of Fermi GBM GRBs' spectra}",
      journal = {\mnras},
         year = 2018,
        month = mar,
       volume = {475},
       number = {1},
        pages = {306-320},
          doi = {10.1093/mnras/stx3152},
archivePrefix = {arXiv},
       eprint = {1712.00389},
 primaryClass = {astro-ph.HE},
       adsurl = {https://ui.adsabs.harvard.edu/abs/2018MNRAS.475..306R}
}

@BOOK{Hacking2013-qv,
  title     = "The emergence of probability",
  author    = "Hacking, Ian",
  publisher = "Cambridge University Press",
  edition   =  2,
  month     =  apr,
  year      =  2013,
  address   = "Cambridge, England"
}

@Manual{survival-package,
    title = {A Package for Survival Analysis in R},
    author = {Terry M Therneau},
    year = {2024},
    note = {R package version 3.7-0},
    url = {https://CRAN.R-project.org/package=survival},
  }

\end{document}